\documentclass[11pt]{article}
\usepackage{amsmath,amsfonts,bm}

\def\eqref#1{equation~\ref{#1}}
\def\1{\bm{1}}

\def\vf{{\bm{f}}}

\def\vk{{\bm{k}}}

\def\vq{{\bm{q}}}

\def\vu{{\bm{u}}}

\def\vx{{\bm{x}}}

\DeclareMathAlphabet{\mathsfit}{\encodingdefault}{\sfdefault}{m}{sl}
\SetMathAlphabet{\mathsfit}{bold}{\encodingdefault}{\sfdefault}{bx}{n}

\usepackage[]{acl}

\usepackage{times}
\usepackage{latexsym}
\usepackage[T1]{fontenc}
\usepackage[utf8]{inputenc}
\usepackage{microtype}

\usepackage{url}
\usepackage{graphicx}  %Required
\usepackage{amsfonts}
\usepackage{CJK}
\usepackage{bm}
\usepackage{mathrsfs}
\usepackage{float}
\usepackage{xcolor,colortbl}
\usepackage{adjustbox}
\usepackage{subfigure}
\usepackage{booktabs,multirow}
\usepackage{bigstrut}
\usepackage{paralist}
\usepackage{bbm}
\usepackage{makecell}
\usepackage{nicefrac}       % compact symbols for 1/2, etc.
\usepackage{microtype} 
\usepackage{epsfig}
\usepackage{diagbox}
\usepackage{framed}
\usepackage{empheq}
\usepackage{array}
\usepackage{enumitem}
\usepackage{mdwlist}
\usepackage{xspace,mfirstuc,tabulary}
\usepackage{tcolorbox}
\usepackage{algorithm}
\usepackage{placeins}
\usepackage{algpseudocode}
\usepackage{microtype}

\usepackage[normalem]{ulem}
\useunder{\uline}{\ul}{}
\usepackage{threeparttable}
\usepackage{makecell}
\usepackage{booktabs}

\usepackage{xspace}
\usepackage{natbib}
\usepackage{hyperref}
\usepackage{url}

\usepackage{graphicx}
\usepackage{subfigure}
\usepackage{multirow}
\usepackage{multicol}
\usepackage{pifont}
\usepackage{enumitem}
\usepackage{amsmath}
\usepackage{appendix}

\usepackage{tikz}
\newcommand*{\circled}[1]{\lower.7ex\hbox{\tikz\draw (0pt, 0pt)%
    circle (.5em) node {\makebox[1em][c]{\small #1}};}}
\usepackage{amsmath}
\usepackage{amssymb}
\usepackage{mathtools}
\usepackage{amsthm}
\usepackage{listings} %
\newcommand{\ie}{{\emph{i.e.,}}\xspace}

\title{HiRoPE: Length Extrapolation for Code Models Using Hierarchical Position}
\author{Kechi Zhang, \ Ge Li\footnotemark[1], \ Huangzhao Zhang, \ Zhi Jin\footnotemark[1] \\
Key Lab of High Confidence Software Technology (PKU), Ministry of Education \\
School of Computer Science, Peking University, China \\
\texttt{\{zhangkechi,lige,zhang\_hz,zhijin\}@pku.edu.cn}}

\begin{document}
\maketitle
\renewcommand{\thefootnote}{\fnsymbol{footnote}}
\footnotetext[1]{Corresponding authors.}
\renewcommand{\thefootnote}{\arabic{footnote}}
\begin{abstract}
Addressing the limitation of context length in large language models for code-related tasks is the primary focus of this paper. Existing LLMs are constrained by their pre-trained context lengths, leading to performance issues in handling long complex code sequences.
Inspired by how human programmers navigate code, we introduce Hierarchical Rotary Position Embedding (HiRoPE), a novel approach that enhances the traditional rotary position embedding into a hierarchical format based on the hierarchical structure of source code. 
HiRoPE offers easy integration into existing LLMs without extra training costs.
Our method is extensively evaluated with various LLMs, demonstrating stable performance in tasks such as language modeling and long code completion. We also introduce a new long code understanding task with real-world code projects, in hopes of promoting further development in this code-related field. 
Theoretically and experimentally, we find that HiRoPE also addresses the out-of-distribution issue in position encoding. Our HiRoPE significantly expands the context length capabilities of LLMs, enabling inference at lengths exponentially greater than the training length. 

% enhancing the traditional rotary position embedding into a hierarchical format based on the hierarchical structure of source code, providing improved extrapolation capabilities.
% This method, inspired by the human approach to code navigation, enhances the traditional rotary position embedding by integrating hierarchical information, enabling more effective handling of long dependencies in code.
% The key contribution of this work is the development of HiRoPE, which not only resolves the long-standing issue of context length limitation in LLMs but also expands their application scope in complex code-related tasks. Our findings show that HiRoPE significantly enhances the context length capability of LLMs, marking a substantial advancement in the field.
\end{abstract}

% \nocite{Ando2005,andrew2007scalable,rasooli-tetrault-2015}

\maketitle

\section{Introduction}

Large language models (LLMs) such as LLaMA-2 \cite{llama2}, and CodeLLaMA \cite{roziere2023code} have achieved significant performances in code-related tasks. These Transformer-based models excel in code comprehension and generation but face a notable challenge: the limitation of maximum context length. LLMs are typically pre-trained with a context length ranging from 2k to 16k tokens, which often proves insufficient for complex, extended source code. Exceeding this length limitation during inference may lead to performance degradation for these code models, particularly in tasks like project-level code completion or long code generation.

Various methods have been developed to extend the context window of LLMs. Some approaches involve fine-tuning on extensive texts \cite{longcontextscale, PI, CLEX, yarn},  which can be resource-intensive and potentially lead to overfitting and loss of performance on shorter sequences. There are also some training-free methods \cite{attnsink, llminf, longnet}. However, these methods usually use window attention rely on local information, and ignore the long dependency in code. It is essential to incorporate \textit{certain structural characteristics} of the code into position encoding to efficiently model these long-distance code dependencies. 

\begin{figure}[!t]
\centering
  \includegraphics[width=0.6\columnwidth]{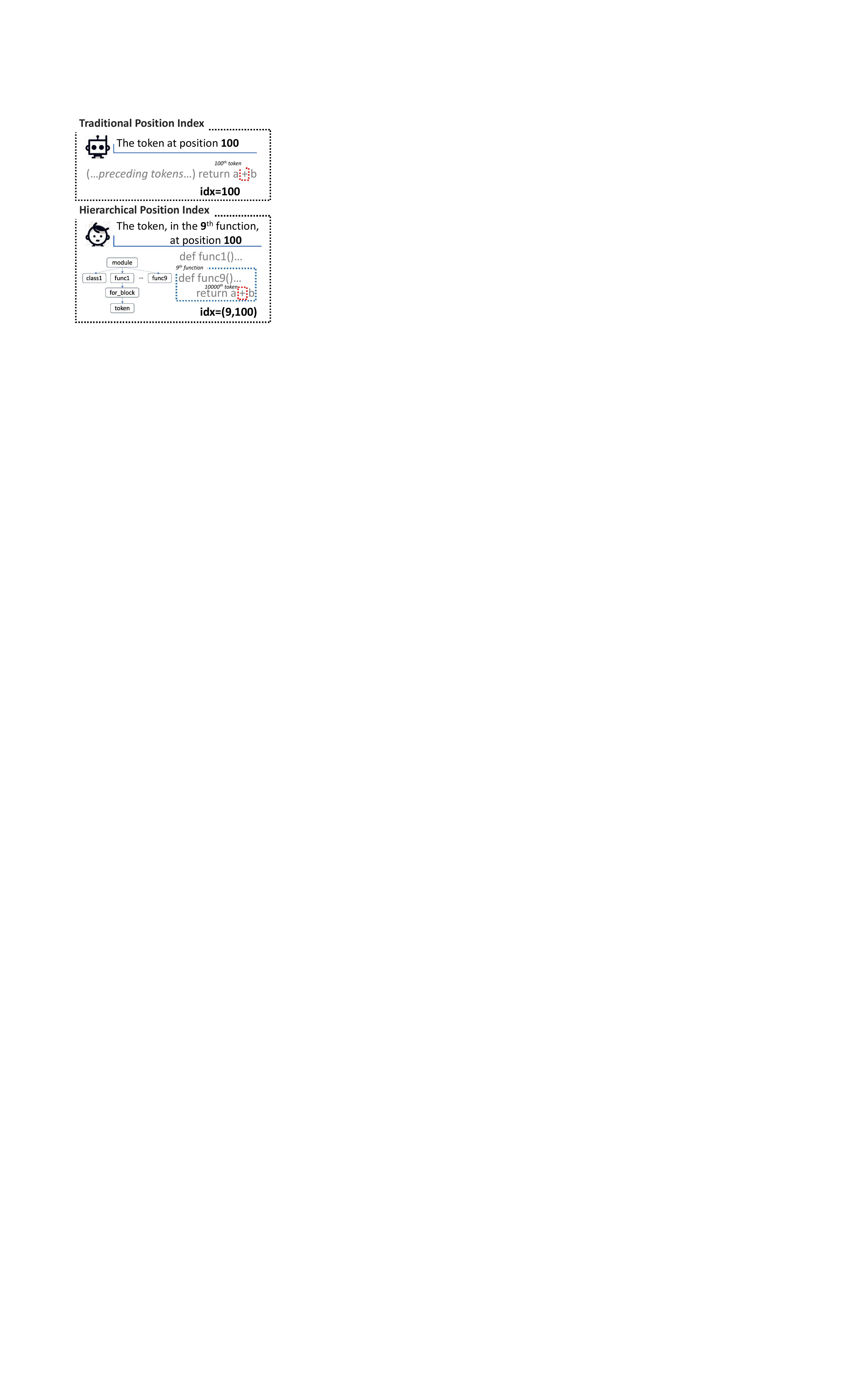}  
\caption{Illustration of the hierarchical position in source code, such as function-level and token-level positions. We also show a simplified abstract syntax tree of the code in the bottom left corner. }
\label{fig:motivation}
% \vspace{-18pt}
\end{figure}

Our work diverges from these methods by focusing on the hierarchical information of source code in position encoding, inspired by how human programmers navigate code. Traditional positional encoding uses token counts for positioning, and treats code as plain text. However, human programmers often use hierarchical information in the code, representing positions in the code efficiently through multi-level hierarchical positions. We propose a hierarchical position approach that identifies token positions within specific levels, such as functions or statements. Figure \ref{fig:motivation} shows the comparison of the traditional position and our hierarchical position. 
It is clear that the hierarchical positional encoding, benefiting from the full utilization of structural information in the code, can more conveniently locate positional information within long code sequences.
This method could more effectively model long dependencies in source code. 

Following such inspirations, we introduce a novel approach, Hierarchical Rotary Position Embedding (HiRoPE), which enhances the popular rotary position embedding (RoPE) \cite{su2021roformer} into a hierarchical format. HiRoPE differentiates itself by extracting hierarchical information from the source code and splitting the RoPE dimension to represent different hierarchical levels. It simultaneously models token-level relative location and higher-level relative location information. We also add a window mechanism to ensure stability with short texts, aligning with traditional positional encoding.

HiRoPE is a plug-and-play solution, easily integrated into existing LLMs without additional training costs. Our extensive experiments with popular LLMs on tasks like language modeling and token completion in long code contexts \cite{CodeParrot} demonstrate its effectiveness. We compare HiRoPE with existing length extrapolation methods using long code benchmarks such as LCC \cite{lcc} and RepoBench \cite{repobench}. We also introduce a new long code understanding task named code symbol understanding with real-world code libraries. 
Theoretically and experimentally, we find that HiRoPE effectively addresses the out-of-distribution issue \cite{ropeood} in position encoding. Our HiRoPE significantly expands the context length capabilities of LLMs, enabling inference at lengths exponentially greater than the training length. 
We believe our work with HiRoPE not only addresses a critical length limitation in LLM applications but also opens new avenues for long-structured data modeling research.

In summary, we make the following main contributions: 
\begin{itemize}
    \item We propose Hierarchical RoPE (HiRoPE), enhancing the traditional rotary position embedding into a hierarchical format based on the hierarchical structure of source code, providing improved extrapolation capabilities.
    \item We conducted comprehensive experiments with LLMs on various long code tasks involving language modeling and code completion. We also introduce a new long code understanding task with real-world code projects, in hopes of promoting further development in this code-related field.
    \item We demonstrate that HiRoPE effectively addresses the out-of-distribution issue in position encoding, enabling inference at lengths exponentially greater than the training length. 
\end{itemize}

\section{Preliminary}
\label{sec:preliminary}
We first introduce rotary position embedding in Transformer in Section \ref{sec:preliminary_rope}. While existing work usually regards source code as plain text for modeling, we will also introduce the ignored hierarchical information in source code in Section \ref{sec:preliminary_hipos}. 

\subsection{Rotary Position Embedding in Transformer}
\label{sec:preliminary_rope}
\def\vf{\mathbf{f}}
\def\vx{\mathbf{x}}
\def\vk{\mathbf{k}}
\def\vq{\mathbf{q}}
\def\vu{\mathbf{u}}
\def\di{\mathrm{i}}

Transformer models require explicit positional information to be injected, typically in the form of positional encodings, to represent the order of inputs. 
Recently, the rotary position embedding (RoPE) \cite{su2021roformer} has become one of the most popular and elegant position encoding strategies and is adopted by various LLMs \cite{Llama, llama2, roziere2023code}. The main point of the RoPE method is using absolute position encoding to show relative position information.
Formally, given a position index $m \in [0, L)$ and an embedding vector $\vx := [x_0, x_1, \ldots, x_{d-1}]^\top$, where $d$ is the dimension of the attention head, RoPE defines a complex function $\vf(\vx, m)$ as follows:
\begin{equation}
\small
\begin{split}
    \vf(\vx,m) = [(x_0 + \di x_1) e^{\di m \theta_0}, (x_2 + \di x_3) e^{\di m \theta_1}, \ldots, \\
    (x_{d-2} + \di x_{d-1})e^{\di m \theta_{d/2-1}}]^\top
\end{split}
\label{eq:fxm}
\end{equation}
where $\di := \sqrt{-1}$ is the imaginary unit and $\theta_j = 10000^{-2j/d}$. 

With RoPE, the self-attention score can be calculated as:

\begin{equation}
\small
\begin{split}
& a(m,n) = \mathrm{Re}\langle\vf(\vq, m), \vf(\vk, n)\rangle  \\
&= \mathrm{Re}\left[\sum_{j=0}^{d/2-1} (q_{2j} +\di q_{2j+1})(k_{2j} - \di k_{2j+1}) e^{\di (m-n)\theta_j}\right]   \\
&= \sum_{j=0}^{d/2-1} [ (q_{2j} k_{2j} + q_{2j+1}k_{2j+1})\cos((m-n)\theta_j) \\
&\quad + (q_{2j} k_{2j+1} - q_{2j+1}k_{2j})\sin((m-n)\theta_j) ] \\
&=: a(m-n)
\end{split}
\label{eq:amn}
\end{equation}
Here $\mathrm{Re}\langle\dots\rangle$ denotes the real part function for a complex number. 
$\vq$ and $\vk$ are the query and key vector for a specific attention head. At each layer, RoPE is applied on both query and key embeddings for computing attention scores. 
We can observe that the calculated attention score is only dependent on relative position $m-n$ through trigonometric functions, which reflects the core of RoPE that uses the absolute position to represent the relative distance. 
Existing studies show that when dealing with long plain text, the RoPE will meet O.O.D issues where the value of $m-n$ during inference is unseen \cite{ropeood}, leading to poor performances.

\subsection{Hierarchical Position in Source Code}
\label{sec:preliminary_hipos}
Most LLMs treat source code as plain text, processing it as if it were ordinary natural language. However, it is essential to take the structural information of code into mind. Source code can be transformed into abstract syntax trees, and these tree structures contain rich hierarchical position information. For example, the code snippets usually can be split into several class or function units, and each class/function contains various types of code blocks and statements. Figure \ref{fig:motivation} shows an illustration of the simplified abstract syntax tree for a code snippet in the bottom left corner. 
This higher-level positional information contains rich semantics of source code, making it easy for human programmers to locate and refer to different semantic parts. Therefore, in many existing program representation tasks, this hierarchical information plays a very important role \cite{codegraph, hit}. However, for today's large language models, this high-level hierarchical positional information is almost ignored. In this paper, we try to incorporate this hierarchical information into the position encoding method.

\section{Hierarchical RoPE}

\begin{figure*}[!t]
\centering
  \includegraphics[width=1.7\columnwidth]{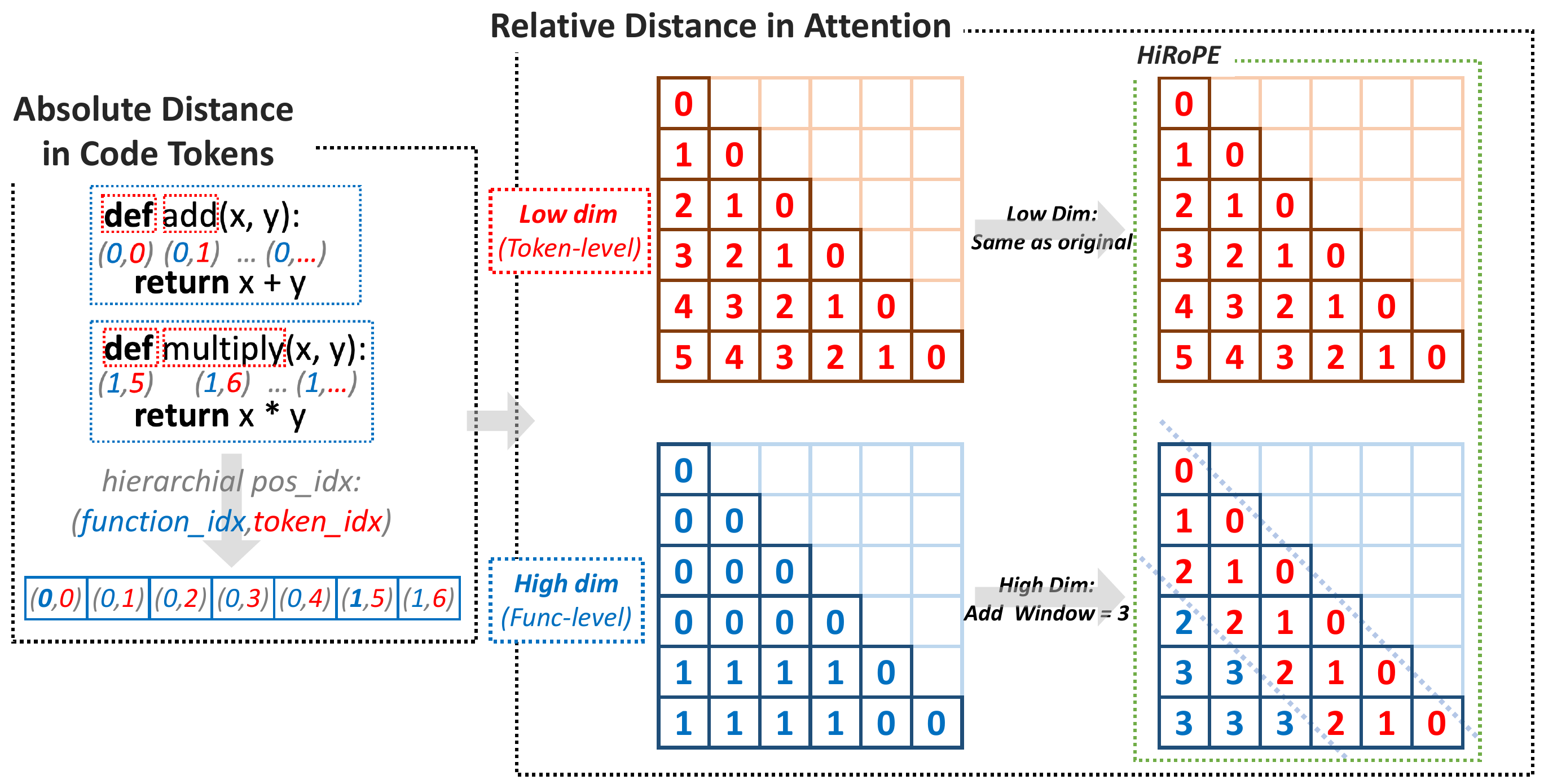}  
\caption{Overview of our HiRoPE. We transfer the existing position encoding method into a hierarchical format (\ie, function-level and token-level) and apply it across different dimensions. We also add a window mechanism to ensure performance stability (in this figure we set $L_{window}$ to 3).}
\label{fig:method}
% \vspace{-18pt}
\end{figure*}

In this paper, we propose HiRoPE, a hierarchical rotary position encoding for source code modeling. 
Our proposed HiRoPE requires two modified stages: \ding{182} \textbf{Hierarchical Format}: We first take the step to transfer the existing rotary position embedding into a hierarchical format. We verify our approach theoretically and find that the hierarchical format can bring stable extrapolation ability for RoPE.
\ding{183} \textbf{Window Mechanism}. To ensure performance stability without further training, we also add a window mechanism, so that when dealing with short texts, our proposed method is consistent with the original positional encoding. An illustration of our HiRoPE is shown in Figure \ref{fig:method}.

\subsection{Hierarchical format}
Unlike previous work that encodes the information of each position as a number index $m \in [0, L)$, we use a h-dimensional vector to represent the hierarchical position index from high-level to low-level, where $h$ is a hyperparameter that indicates how many levels of information we consider in the encoding. 
We begin with a simple case that we set $h = 2$ and each token position can be represented as $(m_1, m_2)$.
We use the higher and lower dimensions in RoPE respectively to represent these two hierarchical position indexes, so the Equation \ref{eq:fxm} can be rewritten as:
\begin{equation}
\small
\begin{split}
   & \vf’(\vx,\textcolor{red}{m_1}, \textcolor{blue}{m_2}) = \\
    & [(x_0 + \di x_1) e^{\di \textcolor{red}{m_1} \theta_0}, \ldots, (x_{d_{s}-2} + \di x_{d_{s}-1})e^{\di \textcolor{red}{m_1} \theta_{d_s/2-1}}, \\
    & (x_{d_{s}} + \di x_{d_{s}+1})e^{\di \textcolor{blue}{m_2} \theta_{d_s/2}}, \ldots, (x_{d-2} + \di x_{d-1})e^{\di \textcolor{blue}{m_2} \theta_{d/2-1}}]^\top
\end{split}
\label{eq:fxm1m2}
\end{equation}
There are a total of $d$ dimensions in RoPE, and we use the lower $d_s$ dimensions to represent the hierarchical index of $m_1$, and the remaining dimensions to represent $m_2$. When we apply it to self-attention, we can get a new calculation of the attention score:
\begin{equation}
\small
\begin{split}
& \mathrm{hierarchicalAttn} = \mathrm{Re}\langle\vf(\vq, \textcolor{red}{m_1}, \textcolor{blue}{m_2}), \vf(\vk, \textcolor{red}{n_1}, \textcolor{blue}{n_2})\rangle  \\
&= \sum_{j=0}^{d_s/2-1} [(q_{2j} k_{2j} + q_{2j+1}k_{2j+1})\cos((\textcolor{red}{m_1-n_1})\theta_j) \\
&\quad + (q_{2j} k_{2j+1} - q_{2j+1}k_{2j})\sin((\textcolor{red}{m_1-n_1})\theta_j)]   \\
&\quad + \sum_{j=d_s/2+1}^{d/2-1} [(q_{2j} k_{2j} + q_{2j+1}k_{2j+1})\cos((\textcolor{blue}{m_2-n_2})\theta_j) \\
&\quad + (q_{2j} k_{2j+1} - q_{2j+1}k_{2j})\sin((\textcolor{blue}{m_2-n_2})\theta_j)] \\
&=: a’(\textcolor{red}{m_1-n_1}, \textcolor{blue}{m_2-n_2})
\end{split}
\label{eq:amn}
\end{equation}
The attention score $a’(\dots)$ we ultimately obtained through the inner product is quite elegant. It includes the relative position distance of various hierarchical levels, and this form can be similarly extended to the representation of more hierarchical positional structures, as shown in Equation \ref{eq:hhiattention}. Specifically, when $h = 1$, it remains the same as the original RoPE.
\begin{equation}
\begin{split}
& \mathrm{hierarchicalAttn_h} \\
&= \mathrm{Re}\langle\vf(\vq, m_{1}, \dots, m_{h}), \vf(\vk, n_{1}, \dots, n_{h}))\rangle  \\
&=: a’(m_{1} - m_{2}, \dots, m_{h} - n_{h})
\end{split}
\label{eq:hhiattention}
\end{equation}

It indicates that the RoPE has the potential to be transformed into a hierarchical form, and we can use the hierarchical position index in the source code for this new form of RoPE, as shown in the left part of Figure \ref{fig:method}.
% , and such format can also be transformed into the computational efficient realization as shown in Appendix \ref{eq:fxmcossin}.

The original RoPE sets $\theta_j$ to $10000^{-2j/d}$, which means that the lower the dimension, the higher its frequency, and the more emphasis on modeling the relative position information of short distances \cite{giraffe}. 
Therefore, in our hierarchical format, we use those low dimensions to represent token-level information, and high dimensions to represent higher-level hierarchical information, such as the function level or statement level in source code.

\subsection{Window Mechanism}
To ensure performance stability without further training, we follow existing extrapolation methods \cite{attnsink, llminf} and add a window mechanism. Specifically, when dealing with short code snippets, we believe that existing LLMs have already mastered the ability to model these short semantic structures from vast pre-training code datasets. 
Therefore, when calculating the attention score, we directly use the original RoPE for those parts that are shorter than a specific length $L_{window}$. And for those long context parts that the distance is larger than $L_{window}$, we transfer them to the new hierarchical format by adding $L_{window} - 1$ to each high-level distance. Our subsequent experiments have proved that even without any additional training, this window mechanism can bring strong stability, making the model's performance applicable in various scenarios with arbitrary input lengths. An illustration of the final HiRoPE is shown in the right part of Figure \ref{fig:method}.
% We give the final python style pseudo codes of our proposed HiRoPE for $h = 2$ in Equation \zkc{add}\ref{alg:hirope}. 

\section{Experiment Setup}

\begin{table}[!t] 
\footnotesize
    \centering
    \resizebox{\linewidth}{!}{
\centering
% Please add the following required packages to your document preamble:
% \usepackage{multirow}

\begin{tabular}{cclcc}
\toprule
Task                                                      & \multicolumn{2}{c}{Dataset}                 & Avg. Length & Samples \\
\midrule
\multirow{4}{*}{\makecell{Long Code \\ Language Modeling}}              & \multirow{4}{*}{CodeParrot}   & 0-2048      &      1031.46       & 100     \\
                                                          &                               & 2048-4096   &            3667.76 & 100     \\
                                                          &                               & 4096-8192   &            7074.57 & 100     \\
                                                          &                               & 8192-16384  &            14353.94 & 100     \\
\midrule
\makecell{Long Text \\ Language Modeling}                               & \multicolumn{2}{c}{ReRoPE-eval}             &     21367.55        & 200     \\
\midrule
\makecell{Code Symbol \\ Understanding}                                 & \multicolumn{2}{c}{Real-world Code Project} &     12976.89        & 56     \\
\midrule
\multicolumn{1}{c}{\multirow{2}{*}{\makecell{Long Code \\ Completion}}} & \multicolumn{2}{c}{LCC}                     &      17855.73       & 300     \\
\multicolumn{1}{l}{}                                      & \multicolumn{2}{c}{RepoBench}               &  21103.42           & 300    \\
\bottomrule
\end{tabular}

}
\caption{Statistics of the evaluation datasets}
   \label{tab:data}
\end{table}

In this section, we aim to answer the following research questions through a series of experiments. 
Details of the evaluation dataset statistics are shown in Table \ref{tab:data}.

\textbf{RQ1.  How is the language modeling capability of HiRoPE on long code sequences?}
We evaluate HiRoPE's language modeling ability on \textit{\textbf{CodeParrot-valid}} dataset \cite{CodeParrot} in Section \ref{sec:longcodeeval}. 
% We divide the original dataset into different length intervals, and for each interval, we select 100 samples for measurement. 
% We use the \textbf{loss}, \textbf{PPL} and \textbf{token completion accuracy} as the metric.

\textbf{RQ2.  How is the language modeling capability of HiRoPE on long natural language sequences? }
The natural language lacks the explicit hierarchical structure information found in code, so we have made some modifications: we set every 128 tokens as a segment, and encode it as higher-level position information. We use the evaluation dataset from \textit{\textbf{ReRoPE-eval}} \footnote{\url{https://github.com/bojone/rerope/blob/main/samples_15k.jsonl}} \cite{rerope} in Section \ref{sec:nleval}. It is a dataset curated from Common Crawl \cite{commoncrawl}, refined by length-based selection criteria.

\textbf{RQ3. How does HiRoPE perform in understanding real-world, long-code projects?} 
To evaluate the effect of the method in real long-code scenarios, we design a new evaluation task on real code projects: \textit{\textbf{Code Symbol Understanding}} in Section \ref{sec:codeprojecteval}. 
Given a long code file, the model is required to output all the function names and class names defined in it. 
We extract long code files from popular open-sourced code repositories, especially those newly updated code projects to avoid data leakage. Details of these code projects are shown in Table \ref{tab:codeprojectstat}.

\textbf{RQ4. How does HiRoPE perform on existing benchmarks for long code completion? } 
We further perform the evaluation using two long code completion benchmarks: \textit{\textbf{LCC}} \cite{lcc} and \textit{\textbf{RepoBench}} \cite{repobench} in Section \ref{sec:lcceval}. 
% These benchmarks are collected from GitHub and the performance can indicate the performance in real-world code completion scenarios.
% These two existing benchmarks contain some code-related tasks and we evaluate our HiRoPE on these tasks. Results are shown in Table \zkc{add a table}.

\textbf{RQ5. What is the impact of various settings in HiRoPE? } 
To demonstrate that each setting in the design of our HiRoPE works, we carry out extensive ablation studies that include the dimensions' split settings, the window mechanism, and the high-level segment split strategy in Section \ref{sec:ablation}.

\subsection{Base LLMs}
\label{sec:setupLLM}

\begin{table}[!t] 
\footnotesize
    \centering
    \resizebox{\linewidth}{!}{
\centering
% Please add the following required packages to your document preamble:
% \usepackage{multirow}

\begin{tabular}{l|cccc}
\toprule
               & LLaMA-2 & ShearedLLaMA & TinyLLaMA & Vicuna \\
\midrule
Para.          & 7B      & 1.3B         & 1.1B      & 7B     \\
\midrule
$L_{pretrain}$   & 4096    & 4096         & 2048      & 2048   \\
\midrule
Vocab Size     & 32000   & 32000        & 32000     & 32000  \\
\midrule
Hidden Size    & 4096    & 2048         & 2048      & 4096   \\
Attention Head & 32      & 16           & 32        & 32     \\
RoPE Dim       & 128     & 128          & 64        & 128   \\
\bottomrule
\end{tabular}

}
\caption{Statistics of base LLMs}
   \label{tab:model}
\end{table}

The models used include LLaMA-2 (7B) \cite{llama2}, Sheared-LLaMA (1.3B) \cite{shearedllama}, TinyLlama (1.1B) \cite{tinyllama}, and Vicuna (7B) \cite{chiang2023vicuna}. This model choice is driven by their widespread use and popularity, as well as the constraints of our computing capabilities.
Details are provided in Table \ref{tab:model}.

\subsection{Baselines}

Considering training on long context sequences is resource-intensive and time-consuming, we focus on those popular length extrapolation methods without training, including NTK \cite{ntk}, ReRoPE \cite{rerope} and Self-Extend \cite{selfextend}.
ReRoPE sets a window when calculating the relative distance, and clips those distances outside the window.
Self-Extend uses group attention with the floor operation to calculate relative distances.
These methods have shown impressive performance on long context language modeling. We also make comparisons with the original RoPE method, which is denoted as ``origin'' in the first row of Table \ref{tab:codeparrot}, \ref{tab:nl} and \ref{tab:codeproject}.

\subsection{Inference Settings}
In our experiments, our HiRoPE uses a two-layer hierarchy, accounting for the position index at the token and function/class levels of the source code based on \textit{tree-sitter} \cite{treesitter}. For long context in natural language, we make some modifications and set every 128 tokens as a higher-level segment.
We set the split dimension half of the total: $d_s  = 0.5 * d_{total}$, and choose a window length: $L_{window} = 512$. We keep the hyperparameters the same for those state-of-the-art baselines for a fair comparison.
We use greedy search decoding for generation. We use 4 A6000 GPUs for all experiments.

\subsection{Details of Code Symbol Understanding task}

In RQ3, we design a new task: Code Symbol Understanding. Given a long code context, the model is
required to output all the function names and class names defined in it. 
We extract long code files from popular open-sourced code repositories, especially those newly updated code projects to avoid data leakage.
We construct a static analysis tool to get the abstract syntax tree of each code file, and then get all defined function and class names in it as the ground-truth output symbols.
Details of these code projects are shown in Table \ref{tab:codeprojectstat}. We show the number of files we extracted from each project as well as the average length of these code files.
We also show the average number of symbols and their location statistics in the table.

\begin{figure}[!h]
\centering
  \includegraphics[width=0.7\columnwidth]{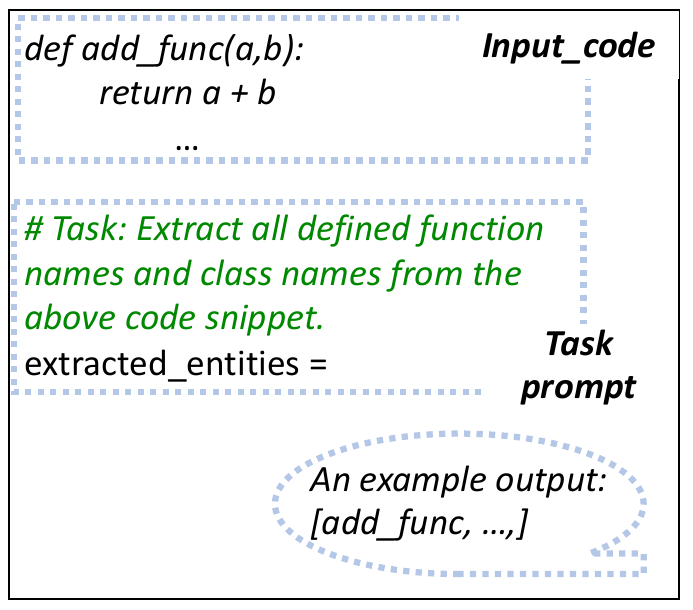}  
\caption{Illustration of Code Symbol Understanding task. We use the task prompt to guide models to extract and output all defined function and class names in input code.}
\label{fig:newtask}
% \vspace{-18pt}
\end{figure}

In Figure \ref{fig:newtask} we show an illustration of our proposed new task. We replace the \textit{input\_code} part with each code file, and use the task prompt to guide models to extract and output all defined function and class names in input code. We also show an example output in the figure.

\begin{table*}[!h] 
\footnotesize
    \centering
    \resizebox{0.8\linewidth}{!}{
\centering
% Please add the following required packages to your document preamble:
% \usepackage{multirow}

\begin{tabular}{l|c|c|c|c|c}
\toprule
Github Repo              & File Nums   & File Length            & Avg. Symbols   & Min. Symbol Loc & Max. Symbol Loc \\
\midrule
ddbourgin/numpy-ml       & 2           & 14055             & 15             & 927             & 10845           \\
gradio-app/gradio        & 1           & 12938             & 11             & 272             & 12265           \\
huggingface/accelerate   & 4           & 12814.5           & 11.5           & 379             & 12500           \\
huggingface/diffusers    & 4           & 13324.75          & 16.2           & 276             & 8347            \\
huggingface/optimum      & 1           & 13491             & 11             & 580             & 12868           \\
huggingface/peft         & 1           & 15457             & 11             & 514             & 8525            \\
huggingface/transformers & 18          & 12919.4           & 12.7           & 239             & 14514           \\
langchain-ai/langchain/  & 2           & 14195             & 11.5           & 285             & 10790           \\
numpy/numpy              & 6           & 11158.8           & 14             & 100             & 12026           \\
tensorflow/tensorflow    & 17          & 13191.7           & 13.4           & 285             & 14920           \\
\midrule
\textbf{Total}           & \textbf{56} & \textbf{12976.89} & \textbf{13.17} & \textbf{100}    & \textbf{14920}  \\
\bottomrule
\end{tabular}

}
\caption{Statistics of Code Symbol Understanding task. We show the number of files we extracted from each project as well as the average length of these code files. We also show the average number of symbols and their location statistics in the table.}
   \label{tab:codeprojectstat}
\end{table*}

\section{Results and Analyses}

\subsection{Long Code Language Modeling}
\label{sec:longcodeeval}
\begin{table*}[!t] 
\footnotesize
    \centering
    
\setlength{\tabcolsep}{3pt}
    \resizebox{\linewidth}{!}{
\centering
\begin{tabular}{l|c|c|l|ccc|ccc|ccc|ccc}
\bottomrule
                              &                       &                       & \multicolumn{1}{c}{Dataset Length:} & \multicolumn{3}{c}{0-2048} & \multicolumn{3}{c}{2048-4096} & \multicolumn{3}{c}{4096-8192} & \multicolumn{3}{c}{8192-16384}  \\
                              \midrule
                              & Para.                 & $L_{pretrain}$   &                                    & loss $\downarrow $    & ppl $\downarrow $      & acc $\uparrow $    & loss $\downarrow $     & ppl $\downarrow $       & acc $\uparrow $     & loss $\downarrow $    & ppl $\downarrow $         & acc $\uparrow $    & loss $\downarrow $    & ppl $\downarrow $          & acc $\uparrow $      \\
                              \midrule
\multirow{5}{*}{LLaMA-2}      & \multirow{5}{*}{7B}   & \multirow{5}{*}{4096} & origin                             & \textbf{0.8579}  & \textbf{2.3583}  & \textbf{0.8065} & 0.9820   & 2.6698   & 0.7551  & nan     & nan        & nan    & nan     & nan         & nan        \\
                              &                       &                       & NTK                                & 1.1107  & 3.0365  & 0.7472 & 1.0469   & 2.8488   & 0.7414  & 0.9858  & 2.6800     & 0.7729 & 1.0181  & 2.7678      & 0.7630         \\
                              &                       &                       & ReRoPE                             & 0.8593  & 2.3615  & 0.8054 & 0.7278   & 2.0705   & 0.8121  & 0.6633  & 1.9411     & 0.8411 & 0.7187  & 2.0518      & 0.8243         \\
                              &                       &                       & Self-Extend                        & 0.8588  & 2.3604  & 0.8055 & 0.7209   & 2.0562   & 0.8147  & 0.6519  & 1.9192     & 0.8441 & 0.6983  & 2.0103      & 0.8290         \\
                              &                       &                       & HiRoPE                             & 0.8586  & 2.3598  & 0.8060 & \textbf{0.7185}   & \textbf{2.0514}   & \textbf{0.8153}  & \textbf{0.6482}  & \textbf{1.9121}     & \textbf{0.8452} & \textbf{0.6821}  & \textbf{1.9780}      & \textbf{0.8332}  \\
                              \midrule
\multirow{5}{*}{ShearedLLaMA} & \multirow{5}{*}{1.3B} & \multirow{5}{*}{4096} & origin                             & \textbf{1.2874}  & \textbf{3.6235}  & \textbf{0.7341} & 1.3103   & 3.7074   & 0.7019  & 3.8381  & 46.4375    & 0.4866 & 5.6958  & 297.6160    & 0.3211         \\
                              &                       &                       & NTK                                & 1.6242  & 5.0744  & 0.6607 & 1.5047   & 4.5029   & 0.6619  & 1.3708  & 3.9383     & 0.7015 & 1.3657  & 3.9183      & 0.6964         \\
                              &                       &                       & ReRoPE                             & 1.2897  & 3.6316  & 0.7332 & 1.0497   & 2.8567   & 0.7560  & 0.9699  & 2.6376     & 0.7846 & 1.0044  & 2.7303      & 0.7716         \\
                              &                       &                       & Self-Extend                        & 1.2892  & 3.6300  & 0.7337 & 1.0428   & 2.8371   & 0.7586  & 0.9568  & 2.6034     & 0.7874 & 0.9804  & 2.6656      & 0.7768         \\
                              &                       &                       & HiRoPE                             & 1.2888  & 3.6285  & 0.7338 & \textbf{1.0382}   & \textbf{2.8242}   & \textbf{0.7600}  & \textbf{0.9514}  & \textbf{2.5894}     & \textbf{0.7885} & \textbf{0.9660}  & \textbf{2.6273}      &\textbf{ 0.7811}  \\
                              \midrule
\multirow{5}{*}{TinyLlama}    & \multirow{5}{*}{1.1B} & \multirow{5}{*}{2048} & origin                             & 1.0788  & 2.9410  & 0.7594 & 4.1732   & 64.9235  & 0.4506  & 6.6603  & 780.8009   & 0.2630 & 7.9938  & 2962.5881   & 0.1682         \\
                              &                       &                       & NTK                                & 1.1837  & 3.2664  & 0.7405 & 1.0952   & 2.9899   & 0.7256  & 0.9719  & 2.6430     & 0.7733 & 1.0021  & 2.7240      & 0.7626         \\
                              &                       &                       & ReRoPE                             & 0.9703  & 2.6388  & 0.7877 & 0.8251   & 2.2821   & 0.7905  & 0.7685  & 2.1565     & 0.8210 & 0.8275  & 2.2877      & 0.8040         \\
                              &                       &                       & Self-Extend                        & \textbf{0.9698}  & \textbf{2.6375}  & 0.7856 & \textbf{0.8123}   & \textbf{2.2530}   & 0.7931  & \textbf{0.7577}  & \textbf{2.1333}     & \textbf{0.8235} & 0.8119  & 2.2521      & 0.8073         \\
                              &                       &                       & HiRoPE                             & 0.9743  & 2.6493  & \textbf{0.7881} & 0.8268   & 2.2861   & \textbf{0.7981}  & 0.7683  & 2.1562     & 0.8208 & \textbf{0.8040}  & \textbf{2.2345}      & \textbf{0.8094}  \\
                              \midrule
\multirow{5}{*}{Vicuna}       & \multirow{5}{*}{7B} & \multirow{5}{*}{2048} & origin                             & 1.1787  & 3.2502  & 0.7551 & 4.6046   & 99.9449  & 0.4473  & 7.7207  & 2254.4730  & 0.2597 & 9.9449  & 20846.3927  & 0.1601         \\
                              &                       &                       & NTK                                & 1.3417  & 3.8255  & 0.7150 & 1.2587   & 3.5208   & 0.7068  & 1.1809  & 3.2573     & 0.7344 & 1.1912  & 3.2911      & 0.7285         \\
                              &                       &                       & ReRoPE                             & 1.0716  & 2.9201  & 0.7800 & 0.8760   & 2.4012   & \textbf{0.7912}  & 0.8138  & 2.2566     & 0.8182 & 0.8580  & 2.3585      & 0.8023         \\
                              &                       &                       & Self-Extend                        & 1.0710  & 2.9183  & \textbf{0.7802} & 0.8735   & 2.3953   & 0.7891  & \textbf{0.7988}  & \textbf{2.2228}     & \textbf{0.8220} & 0.8351  & 2.3049      & 0.8066         \\
                              &                       &                       & HiRoPE                             & \textbf{1.0707}  & \textbf{2.9174}  & 0.7799 & \textbf{0.8724}   & \textbf{2.3926}   & 0.7903  & 0.8002  & 2.2261     & 0.8213 & \textbf{0.8314}  & \textbf{2.2965}      & \textbf{0.8080} \\
                              \bottomrule
                              \end{tabular}
}

\caption{Language Modeling Ability on CodeParrot-valid dataset. ``nan'' indicates that the model performs significantly poor on the given setting.}
\label{tab:codeparrot}
\end{table*}
Language modeling is the most fundamental and the least requirement for a LLM. We evaluate HiRoPE's language modeling ability on CodeParrot-valid dataset. We divide the original dataset into different length intervals. The experiment results are shown in Table \ref{tab:codeparrot}.
A smaller \textit{loss} and a smaller \textit{ppl} indicate a stronger language modeling capacity of the corresponding model. Conversely, a larger \textit{acc} suggests a stronger capability of the respective model in code completion on the given dataset. The \textit{origin} indicates that we directly use the original setting of the model to evaluate. 

Experiments show that original LLMs perform badly on the long code language modeling task. Even when the length slightly exceeds the pre-training length, the ppl of all models exceed 45, demonstrating their essential lack of modeling and understanding capabilities for longer codes. 
When we apply length extrapolation methods, all methods can reduce the loss and ppl into an acceptable range for those long source codes. Specifically, for ultra-long code sequences (length over 8192), HiRoPE achieves the best results in all metrics and under all settings. This fully demonstrates the advantages of our HiRoPE in modeling long sequence codes. Our method also shows generalization abilities. The four models evaluated have differences in model parameters, pre-training data, and pre-training length, yet our method has shown very good results on all these models. 

It is worth noting that our method does not impair the model's performance on shorter code. We noted that some popular length extension methods, such as NTK, can impair the performance on short code. Thanks to our window mechanism, our method stays on par with the baseline model on shorter datasets (length 0-2048) and even surpasses the baseline on some metrics. HiRoPE demonstrates consistently excellent language modeling capabilities in various code length scenarios.

\subsection{Long Text Language Modeling}
\label{sec:nleval}
\begin{table}[!t] 
\footnotesize
    \centering
    
\setlength{\tabcolsep}{1pt} % Default value: 6pt
    \resizebox{\linewidth}{!}{
\centering
% Please add the following required packages to your document preamble:
% \usepackage{multirow}
\begin{tabular}{l|l|ccc|ccc}
\toprule
                              &             & last loss $\downarrow $ & last ppl $\downarrow $   & last acc $\uparrow $ & all loss $\downarrow $ & all ppl $\downarrow $  & all acc $\uparrow $        \\
                              \midrule
\multirow{5}{*}{TinyLlama}    & origin      & 9.3083    & >1000 & 0.0194   & 7.3486   & >1000 & 0.1466          \\
                              & NTK         & 2.1338    & 8.4468     & 0.5553   & 2.2745   & 9.7234    & 0.53            \\
                              & ReRoPE      & 1.8448    & 6.327      & 0.6008   & 1.903    & 6.7062    & 0.5867          \\
                              & Self-Extend & 1.7904    & 5.9916     & 0.6089   & 1.8749   & 6.5201    & 0.5908          \\
                              & HiRoPE      & \textbf{1.7829}    & \textbf{5.9474}     & \textbf{0.6102}   & \textbf{1.8717}   & \textbf{6.4991}    & \textbf{0.5913} \\
                              \midrule
\multirow{5}{*}{ShearedLLaMA} & origin      & 8.5027    & >1000  & 0.0412   & 5.5237   & 250.5  & 0.274           \\
                              & NTK         & 2.3596    & 10.5863    & 0.5159   & 2.4056   & 11.0853   & 0.5059          \\
                              & ReRoPE      & 1.7952    & 6.0205     & 0.605    & 1.8514   & 6.3687    & 0.5932          \\
                              &  Self-Extend     & 1.7662    & 5.8486     & 0.6105   & 1.8348   & 6.264     & \textbf{0.5966} \\
                              & HiRoPE & \textbf{1.7622}    & \textbf{5.8252}     & \textbf{0.6113}   & \textbf{1.8332}   & \textbf{6.2536}    & 0.5963          \\
                              \bottomrule
\end{tabular}

}
\caption{Language Modeling Ability on ReRoPE-eval dataset. In addition to calculating metrics on all tokens (refer to "all\_..."), we also record metrics on the last 2048 tokens of each data (refer to "last\_..." ).}
   \label{tab:nl}
\end{table}

In addition to testing the ability of language modeling on long code, we also evaluate its effects on long natural language texts. We set every 128 tokens as a high-level segment.
We use the ReRoPE-eval dataset, and results are shown in Table \ref{tab:nl}. In addition to calculating metrics on all tokens (refer to "all\_..." in the table), we also record metrics on the last 2048 tokens of each data (refer to "last\_...").

We find that our HiRoPE can also achieve significant improvement. HiRoPE achieves the best results on almost all metrics. 
Another interesting observation is that, given sufficient context, the model can utilize this contextual information to perform better when generating later tokens. That is to say, the metrics of "last\_..." should be better than those of "all\_...". However, we observe that for the original model, the situation is contrary to this.
We attribute this to the fact that the original model's ability to model long sequence languages is so poor that it can't utilize that distant contextual information at all. Our HiRoPE can significantly improve the model's ability to handle long codes and textual data, without requiring any training,  reflecting its practical value.

\subsection{Code Symbol Understanding}
\label{sec:codeprojecteval}
\begin{table}[!t] 
\footnotesize
    \centering
    
\setlength{\tabcolsep}{2pt}
    \resizebox{\linewidth}{!}{
\centering
% Please add the following required packages to your document preamble:
% \usepackage{multirow}

\begin{tabular}{l|cccccccc}
\toprule
                              &        & \multirow{2}{*}{\makecell{Code Symbol \\ Understanding \\ \textbf{\textit{(Recall $\uparrow$)}}}} & \multicolumn{6}{c}{\makecell{Long Code \\ Completion \\ \textbf{\textit{(Edit Sim $\uparrow$)}}}}                     \\ \cline{4-9}
                              &        &                                            & \multicolumn{3}{c}{LCC}      & \multicolumn{3}{c}{RepoBench} \\
                              \midrule
                              &        &                                            & 0-4k     & 4k-8k    & >8k    & 0-4k     & 4k-8k    & >8k     \\
                              \midrule
                              % &        &                                            &          &          &        &          &          &         \\
\multirow{3}{*}{LLaMA-2}      & origin & 0.0012                                     & 54.5     & 4.36     & 4.08   & 8.29     & 6.79     & 6.59    \\
                              & ReRoPE & 0.0837                                     & 65.83    & 67.43    & 63.22  & \textbf{52.82}    & 47.85    & 45.38   \\
                              & HiRoPE & \textbf{0.0911}                                     & \textbf{66.61}    & \textbf{69.93}    & \textbf{65.38}  & 52.20     & \textbf{53.30}     & \textbf{51.24}   \\ \midrule
\multirow{3}{*}{ShearedLLaMA} & origin & 0.0067                                     & 27.17    & 3.33     & 2.52   & 4.39     & 2.94     & 2.35    \\
                              & ReRoPE & 0.0743                                     & 35.56    & 36.1     & 36.91  & 34.03    & 37.37    & 33.44   \\
                              & HiRoPE & \textbf{0.0809}                                     & \textbf{46.13}    & \textbf{51.67}    & \textbf{46.33}  & \textbf{40.17}    & \textbf{39.98}    & \textbf{39.52}   \\ \midrule
\multirow{3}{*}{TinyLLaMA}    & origin & 0.0067                                     & 17.29    & 5.45     & 6.28   & 7.91     & 7.53     & 7.07    \\
                              & ReRoPE & 0.1214                                     & \textbf{49.22}    & \textbf{57.20}     & \textbf{53.11}  & 37.72    & 40.50     & 39.38   \\
                              & HiRoPE & \textbf{0.1415}                                     & 35.17    & 42.83    & 49.92  & \textbf{42.48}    & \textbf{43.53}    & \textbf{39.82}   \\ \midrule
\multirow{3}{*}{Vicuna}       & origin & 0.0067                                     & 18.47    & 2.56     & 2.76   & 3.67     & 2.49     & 2.32    \\
                              & ReRoPE & 0.0636                                     & 57.95    & 59.73    & 58.52  & \textbf{42.78}    & \textbf{43.65}    & 45.23   \\
                              & HiRoPE & \textbf{0.0721}                                     & \textbf{63.42}    & \textbf{62.01}    & \textbf{64.42}  & 37.10     & 42.30     & \textbf{45.93}   \\ \bottomrule
\end{tabular}

}
\caption{Performance on Code Symbol Understanding and Long Code Completion.}
   \label{tab:codeproject}
\end{table}

To evaluate the effect of the method in real long-code scenarios, we have designed an evaluation task for real code projects: Code Symbol Understanding. Given a long code context, the model is required to output all the function names and class names defined in it. These pre-defined functions and classes are reused frequently in actual code development. For code models, understanding which functions and classes are defined in the code project is a basic capability requirement. This task is inspired by the popular "Needle in a Haystack" synthetic evaluation \cite{needle}, but our code symbol understanding task is more realistic and code-related.
Task examples are shown in Figure \ref{fig:newtask}.  We use \textit{recall} as the evaluation metric.

Our experiments in Table \ref{tab:codeproject} have proven that this seemingly simple task is extremely difficult for LLMs. We also evaluate this task using \textit{GPT-3.5-16k} \cite{GPT-3.5} and find that its result is only 0.72. All these LLMs are not ideal in this more realistic code symbol understanding task.
Our HiRoPE has been improved from the perspective of positional encoding, enabling the model to perceive structural hierarchy changes in the code, thus achieving relatively good results. Compared to the original models, our HiRoPE can achieve almost a hundredfold improvement on average across four models.
We release our dataset in hopes of promoting further development in this code-related field.

\subsection{Long Code Completion}
\label{sec:lcceval}
We further perform the evaluation using two real-world long code completion benchmarks: LCC and RepoBench. Given a long code context, the model is required to generate the complete next line of code. We follow the experiment settings in Longbench-E \cite{longbench} and use \textit{edit similarity} as metrics. Results are shown in Table \ref{tab:codeproject}.

Our HiRoPE also achieves stable improvements on this long code-related task. The input code context is filled with various predefined functions and classes. Our method can effectively sense these contents, handle these complex dependencies, and successfully use those functions that are defined far away during generation. Under two dataset scenarios, our method consistently outperforms other baseline settings.  The results reflect the practicality and generalization ability of our method.

\subsection{Ablation Study}
\label{sec:ablation}
\begin{figure}[!t]
\centering
  \includegraphics[width=\columnwidth]{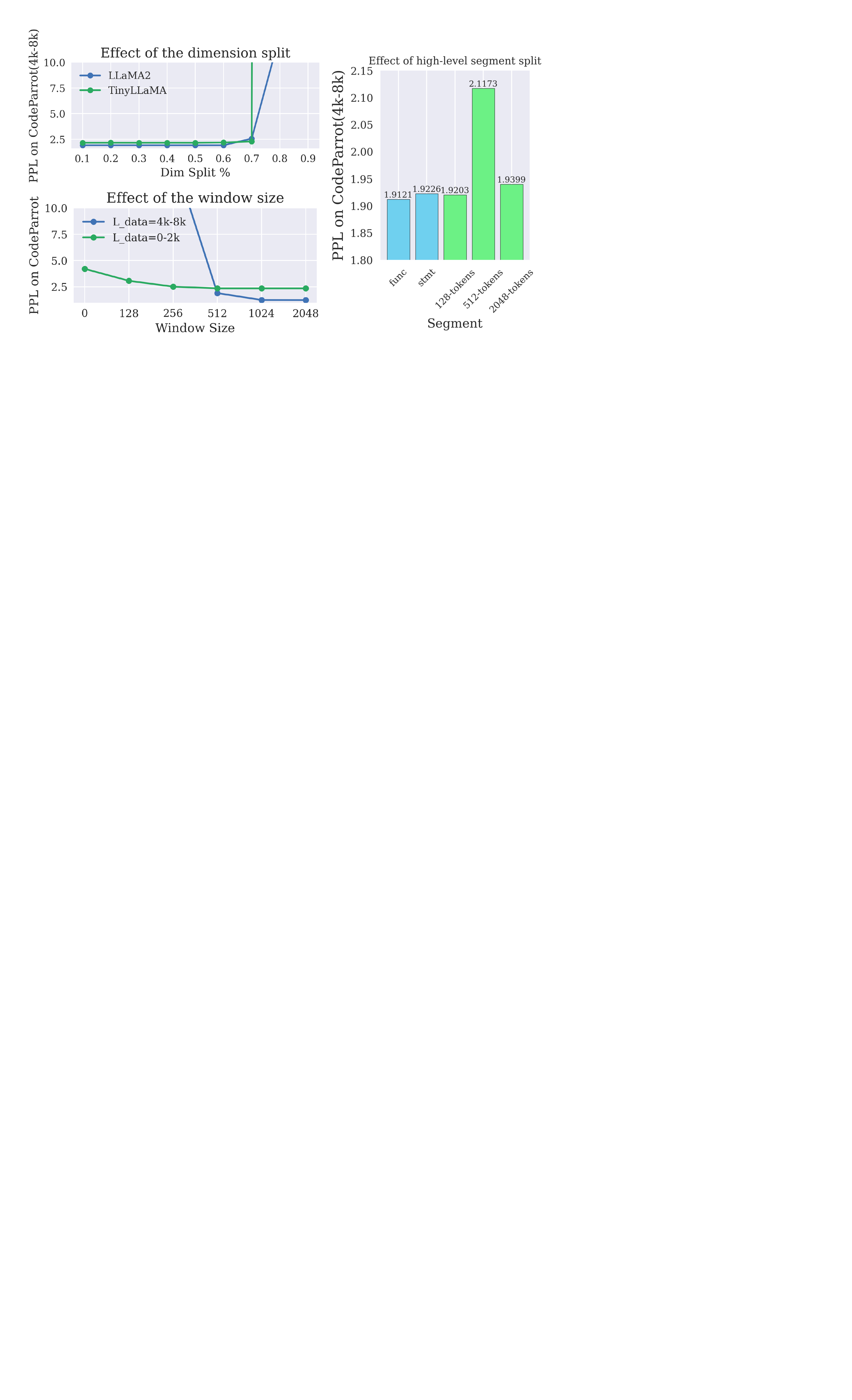}  
\caption{Ablation Studies including the settings of the dimension split, the window mechanism, and the high-level segment split strategy.}
\label{fig:ablation}
% \vspace{-10pt}
\end{figure}
In our experiments, \ding{182} we set the split dimension half of the total $d_s  = 0.5 * d_{total}$ and \ding{182} choose a window length \ding{183} $L_{window} = 512$. \ding{184} Our hierarchical position includes both the token and function/class levels of the source code. We further carry out extensive ablation studies including these settings and results are shown in Figure \ref{fig:ablation}.

We choose different dimension splitting ratios to observe the performance of LLaMA-2 and TinyLLaMA in terms of ppl on CodeParrot [4k-8k]. When the split ratio is 1, the HiRoPE degenerates into the original model. 
Specifically, both models show significant fluctuations in ppl between ratios of $[0.6,0.7]$. We will explore the reason in Section \ref{sec:ood}.

We change the window length to observe the performance of LLaMA-2 on CodeParrot [0-2k] and [4k-8k]. For shorter code data, we observe a decreasing trend in ppl as the window size increases. This validates that the window mechanism can allow the model to retain its original computational mechanisms and better handle short-distance dependencies. For longer code data, ppl behaves anomalously when the window size is very small. This also indicates that window mechanisms play a key role in modeling long-distance dependencies.

We change the high-level segment split strategy. In addition to dividing the hierarchy at the func-level, we also try to split at the code statement level as well as implementing a strategy of splitting continuous n-tokens as a high-level segment (n = 128, 512, 1024). Experiments show that dividing at the function level achieves the best results. The semantics within a function are relatively similar, while the semantics between functions usually vary greatly. It is necessary to divide long code sequences into levels according to functions and classes.

\section{Discussion}

\subsection{Mitigating Out-of-Domain Issues in Long Contexts}
\label{sec:ood}

Existing work \cite{ropeood} shows that LLMs fail on input out of the pretraining context window because of the Out-Of-Distribution (O.O.D) issues.
We take the inspiration to explain it from a cyclical perspective. 
In RoPE, each dimension is composed of trigonometric functions, and its period can be denoted as $T_j = \frac{2\pi}{\theta_j} = 2\pi * 10000^{\frac{2j}{d}}$. Only those dimensions that have been completely trained within the pre-training length can be considered as a reliable part for extrapolation, others would encounter O.O.D issues when dealing with problems of extrapolation. We can then get those reliable dimensions by calculating $T_j < L_{pretrain}$. The calculated dim split is 0.70 and 0.63 for the four models in our experiments (Table \ref{tab:OOD}). We are surprised to find that it is similar to the ratio we obtained in the ablation study in Figure \ref{fig:ablation}. Our HiRoPE uses those high dimensions to represent higher-level position index information, and properly applies them to smaller input numbers, thus mitigating O.O.D issues in long code contexts.

The traditional RoPE uses a number $m$ as the position index to represent position information. Due to the O.O.D problem in high dimensions, its reliable range is $ \{ m \in [0, L_{pretrain}]\}$. 
In our HiRoPE, we use a two-layer hierarchy as $(m_1, m_2)$ and the reliable range is $\{m_1 \in [0, L_{pretrain}], m_2 \in [0, L_{pretrain}]\}$. It proves that under ideal circumstances, HiRoPE can effectively extrapolate to the length of $L^h$ in an exponential ratio, where $h$ is the hierarchy layer. 
We attempt to explore the upper limit of our HiRoPE's extrapolation performance in the experiment in the next Section \ref{sec:uplimit}.

\subsection{Upper Limit of HiRoPE's Performance}
\label{sec:uplimit}
\begin{figure}[!t]
\centering
  \includegraphics[width=0.7\linewidth]{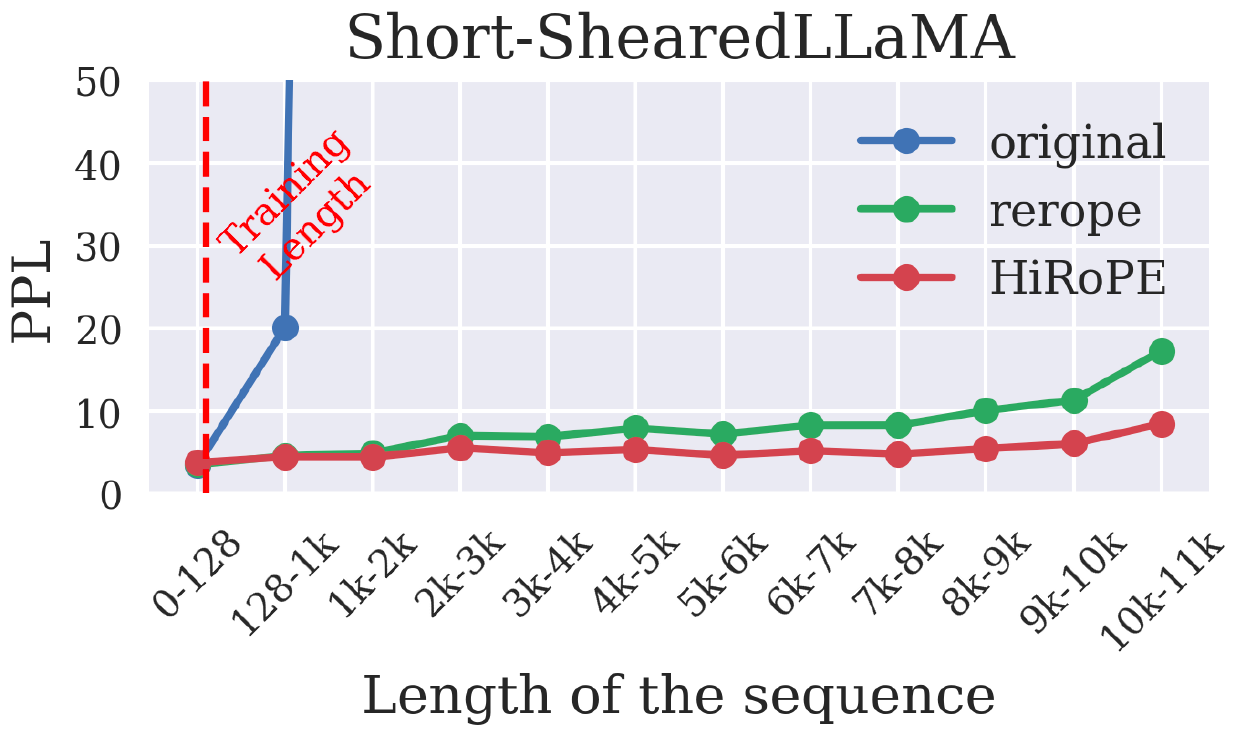}  
\caption{Performance of Short-ShearedLLaMA on CodeParrot dataset. The training length is set to 128. The results suggest our method has the potential to extrapolate code models at an exponential length.}
\label{fig:shortenllm_one}
% \vspace{-10pt}
\end{figure}

% Ideally, for a model of length $L$ during  pre-training, the hierarchical RoPE of $h$ layers can be extended to an outreach length of $L^h$ at the inference stage.
% We try to explore its true performance limit through a series of constructive experiments.

Due to computational resource constraints, we made the following modifications based on Section \ref{sec:longcodeeval}: 
\ding{182} Firstly, in order to obtain a base model of suitable length, we designed a training strategy to obtain a shorter context length LLM, named \textbf{ShortLLM}: We use the position interpolation \cite{PI} method \textbf{\textit{reversely}} to reduce the input length of some mainstream models to $L_{short}=128$ at a smaller training cost. Specifically, given a position index $m$ in the original RoPE, we use the new index $\alpha_{short} * m$ to replace it as shown in Table \ref{tab:shortllmstat}. We fine-tune short models for 1000 steps.
\ding{183} We resample the CodeParrot-valid dataset, further refining it into smaller distance ranges, each range containing up to 50 test samples. 
The results are shown in Figure \ref{fig:shortenllm_one} and \ref{fig:shortenllm}.
% in Table \ref{tab:shortenLLM} and 
% , and the specific training details are displayed in the appendix \zkc{add}.

Our trained ShortLLM successfully demonstrates the expected performance: the performance drastically decreases after surpassing the training length $L_{short} = 128$. We then apply our HiRoPE as well as the baseline ReRoPE. Our HiRoPE demonstrates a more stable trend on long code sequences, even at the position close to $L_{short}^2$. It suggests our method has the potential to extrapolate code models \textbf{at an exponential length}.

\section{Related Work}

Existing large language models are originally trained with fixed context sizes. When dealing with longer sequences, the model's performance may decrease quite drastically.
Recent studies have explored ways to expand the context length. Position Interpolation \cite{PI} linearly down-scales the input position indices to match the original context window size of LLMs with several training steps. Similar studies \cite{CLEX, yarn, ropebucket, lcc} also require fine-tuning. However, these methods need additional tuning in longer contexts or face a disastrous collapse after the extrapolation bound.
There are also some approaches without training. Some studies use window attention to clip the long sequences such as \cite{attnsink, llminf, longnet}). However, these methods rely on local information and may not effectively expand the context window, struggling with long dependencies in code. Recently, some methods have explored modifying the relative distance to extend the extrapolation length \cite{ntk,rerope, selfextend} and focus on the natural language text. 
We pursue the research line of training-free methods and propose considering the structural information of the code when modeling the position.
Our work is inspired by the hierarchical embedding in our former work \cite{hit}. We notice hierarchical positions are used in some concurrent work \cite{twostones,fit} in NLP and CV domains.
We expand the traditional RoPE method into a hierarchical format and prove the effectiveness of our HiRoPE through theoretical derivations and practical experiments.

\section{Conclusion}
We propose HiRoPE, a training-free solution to the context length limitation in LLMs for long code modeling. We integrate the hierarchical structure of source code into position encoding of LLMs. 
Experiments demonstrate that HiRoPE achieves stable improvements on diverse code-related tasks. Our work not only addresses a critical limitation in LLM applications but also opens new avenues for long structured data modeling research.

% Considering the scarcity of computational resources in most usage scenarios, we have focused our experiments on the current untrained version. 
% Our theoretical derivations demonstrate that our method aligns well with the theoretical formulations of the original RoPE, indicating its potential for further training and application in long contexts.
% In the future, we will explore the impact of hierarchical positions with training.

\section{Acknowledgments}

This research is supported by the National Key R\&D Program under Grant No. 2023YFB4503801, the National Natural Science Foundation of China under Grant No. 62072007, 62192733, 61832009, 62192730, and the Major Program (JD) of Hubei Province (No.2023BAA024).

\section*{Limitations}
There are several limitations to our work that we aim to address:

Firstly, constrained by computational resources, we choose models below 7B for experiments. The four models evaluated have differences in model parameters, pre-training data, and pre-training length, yet our method has shown very good results on all these models. We will attempt to conduct experiments on models with larger parameters and more complex structures to promote the development of the LLM community.

Next, our discussion on the upper limit of the HiRoPE's performance tends to lean towards theoretical derivation. We have designed a set of ShortLLM experiments to prove our conclusions. It suggests our method has the potential to extrapolate code models at an exponential length, so for the LLaMA-2 model with $L_{pretrain} = 4096$,  we can \textbf{theoretically} extrapolate its length to  $L_{pretrain}^2 \approx 16,000,000$. We are not clear whether some settings will implicitly affect the performance of the model. We will continue to explore the robustness of this experimental idea and try to explore the maximum performance of our method on real LLMs.

% 限制 acl 2024要求
% 作者必须在专门的“局限性”部分讨论其工作的局限性。这部分应该包含在论文的末尾，在参考文献之前，它不会计入页数限制。这包括长论文和短论文。注意，在2023年12月之前，这是可选的。
%对于ARR提交，参考文献和附录应包含在论文的pdf中，但不计入页数限制。
%正文8页

% \bibliography{main}
% \bibliographystyle{acl_natbib}
% Entries for the entire Anthology, followed by custom entries
\bibliography{main}

\begin{thebibliography}{32}
\expandafter\ifx\csname natexlab\endcsname\relax\def\natexlab#1{#1}\fi

\bibitem[{Allamanis et~al.(2018)Allamanis, Brockschmidt, and Khademi}]{codegraph}
Miltiadis Allamanis, Marc Brockschmidt, and Mahmoud Khademi. 2018.
\newblock \href {https://openreview.net/forum?id=BJOFETxR-} {Learning to represent programs with graphs}.
\newblock In \emph{6th International Conference on Learning Representations, {ICLR} 2018, Vancouver, BC, Canada, April 30 - May 3, 2018, Conference Track Proceedings}.

\bibitem[{Bai et~al.(2023)Bai, Lv, Zhang, Lyu, Tang, Huang, Du, Liu, Zeng, Hou, Dong, Tang, and Li}]{longbench}
Yushi Bai, Xin Lv, Jiajie Zhang, Hongchang Lyu, Jiankai Tang, Zhidian Huang, Zhengxiao Du, Xiao Liu, Aohan Zeng, Lei Hou, Yuxiao Dong, Jie Tang, and Juanzi Li. 2023.
\newblock \href {https://doi.org/10.48550/ARXIV.2308.14508} {Longbench: {A} bilingual, multitask benchmark for long context understanding}.
\newblock \emph{CoRR}, abs/2308.14508.

\bibitem[{bloc97(2023)}]{ntk}
bloc97. 2023.
\newblock Ntk-aware scaled rope allows llama models to have extended (8k+) context size without any fine-tuning and minimal perplexity degradation.
\newblock \url{https://www.reddit.com/r/LocalLLaMA/comments/14lz7j5/ntkaware_scaled_rope_allows_llama_models_to_have/}.

\bibitem[{Brunsfeld et~al.(2024)Brunsfeld, Hlynskyi, Qureshi, Thomson, Vera, and et~al.}]{treesitter}
Max Brunsfeld, Andrew Hlynskyi, Amaan Qureshi, Patrick Thomson, Josh Vera, and et~al. 2024.
\newblock \href {https://doi.org/10.5281/zenodo.10638807} {tree-sitter/tree-sitter: v0.21.0-pre-release-1}.

\bibitem[{Chen et~al.(2023{\natexlab{a}})Chen, Li, Meng, Liang, and Bing}]{CLEX}
Guanzheng Chen, Xin Li, Zaiqiao Meng, Shangsong Liang, and Lidong Bing. 2023{\natexlab{a}}.
\newblock \href {https://doi.org/10.48550/ARXIV.2310.16450} {{CLEX:} continuous length extrapolation for large language models}.
\newblock \emph{CoRR}, abs/2310.16450.

\bibitem[{Chen et~al.(2023{\natexlab{b}})Chen, Wong, Chen, and Tian}]{PI}
Shouyuan Chen, Sherman Wong, Liangjian Chen, and Yuandong Tian. 2023{\natexlab{b}}.
\newblock \href {https://doi.org/10.48550/ARXIV.2306.15595} {Extending context window of large language models via positional interpolation}.
\newblock \emph{CoRR}, abs/2306.15595.

\bibitem[{Chen et~al.(2023{\natexlab{c}})Chen, Lv, Lin, Chen, Wu, Huang, Li, and Yan}]{ropebucket}
Yuhan Chen, Ang Lv, Ting{-}En Lin, Changyu Chen, Yuchuan Wu, Fei Huang, Yongbin Li, and Rui Yan. 2023{\natexlab{c}}.
\newblock \href {https://doi.org/10.48550/ARXIV.2312.04455} {Fortify the shortest stave in attention: Enhancing context awareness of large language models for effective tool use}.
\newblock \emph{CoRR}, abs/2312.04455.

\bibitem[{Chiang et~al.(2023)Chiang, Li, Lin, Sheng, Wu, Zhang, Zheng, Zhuang, Zhuang, Gonzalez et~al.}]{chiang2023vicuna}
Wei-Lin Chiang, Zhuohan Li, Zi~Lin, Ying Sheng, Zhanghao Wu, Hao Zhang, Lianmin Zheng, Siyuan Zhuang, Yonghao Zhuang, Joseph~E Gonzalez, et~al. 2023.
\newblock Vicuna: An open-source chatbot impressing gpt-4 with 90\%* chatgpt quality.
\newblock \emph{See https://vicuna. lmsys. org (accessed 14 April 2023)}.

\bibitem[{CodeParrot(2022)}]{CodeParrot}
CodeParrot. 2022.
\newblock https://huggingface.co/codeparrot.

\bibitem[{Crawl(2023)}]{commoncrawl}
Common Crawl. 2023.
\newblock Common crawl.
\newblock \url{https://commoncrawl.org/}.

\bibitem[{Ding et~al.(2023)Ding, Ma, Dong, Zhang, Huang, Wang, Zheng, and Wei}]{longnet}
Jiayu Ding, Shuming Ma, Li~Dong, Xingxing Zhang, Shaohan Huang, Wenhui Wang, Nanning Zheng, and Furu Wei. 2023.
\newblock \href {https://doi.org/10.48550/ARXIV.2307.02486} {Longnet: Scaling transformers to 1, 000, 000, 000 tokens}.
\newblock \emph{CoRR}, abs/2307.02486.

\bibitem[{gkamradt(2023)}]{needle}
gkamradt. 2023.
\newblock Needle in a haystack - pressure testing llms.
\newblock \url{https://github.com/gkamradt/LLMTest_NeedleInAHaystack/tree/main}.

\bibitem[{GPT-3.5(2023)}]{GPT-3.5}
GPT-3.5. 2023.
\newblock \url{https://platform.openai.com/docs/models/gpt-3-5}.

\bibitem[{Guo et~al.(2023)Guo, Xu, Duan, Yin, and McAuley}]{lcc}
Daya Guo, Canwen Xu, Nan Duan, Jian Yin, and Julian~J. McAuley. 2023.
\newblock \href {https://proceedings.mlr.press/v202/guo23j.html} {Longcoder: {A} long-range pre-trained language model for code completion}.
\newblock In \emph{International Conference on Machine Learning, {ICML} 2023, 23-29 July 2023, Honolulu, Hawaii, {USA}}, pages 12098--12107.

\bibitem[{Han et~al.(2023)Han, Wang, Xiong, Chen, Ji, and Wang}]{llminf}
Chi Han, Qifan Wang, Wenhan Xiong, Yu~Chen, Heng Ji, and Sinong Wang. 2023.
\newblock \href {https://doi.org/10.48550/ARXIV.2308.16137} {Lm-infinite: Simple on-the-fly length generalization for large language models}.
\newblock \emph{CoRR}, abs/2308.16137.

\bibitem[{He et~al.(2024)He, Feng, Luo, Yang, He, Xu, Zhang, Yang, and Wang}]{twostones}
Zhenyu He, Guhao Feng, Shengjie Luo, Kai Yang, Di~He, Jingjing Xu, Zhi Zhang, Hongxia Yang, and Liwei Wang. 2024.
\newblock \href {https://doi.org/10.48550/ARXIV.2401.16421} {Two stones hit one bird: Bilevel positional encoding for better length extrapolation}.
\newblock \emph{CoRR}, abs/2401.16421.

\bibitem[{Jin et~al.(2024)Jin, Han, Yang, Jiang, Liu, Chang, Chen, and Hu}]{selfextend}
Hongye Jin, Xiaotian Han, Jingfeng Yang, Zhimeng Jiang, Zirui Liu, Chia{-}Yuan Chang, Huiyuan Chen, and Xia Hu. 2024.
\newblock \href {https://doi.org/10.48550/ARXIV.2401.01325} {{LLM} maybe longlm: Self-extend {LLM} context window without tuning}.
\newblock \emph{CoRR}, abs/2401.01325.

\bibitem[{Liu et~al.(2023{\natexlab{a}})Liu, Xu, and McAuley}]{repobench}
Tianyang Liu, Canwen Xu, and Julian~J. McAuley. 2023{\natexlab{a}}.
\newblock \href {https://doi.org/10.48550/ARXIV.2306.03091} {Repobench: Benchmarking repository-level code auto-completion systems}.
\newblock \emph{CoRR}, abs/2306.03091.

\bibitem[{Liu et~al.(2023{\natexlab{b}})Liu, Yan, Zhang, An, Qiu, and Lin}]{ropeood}
Xiaoran Liu, Hang Yan, Shuo Zhang, Chenxin An, Xipeng Qiu, and Dahua Lin. 2023{\natexlab{b}}.
\newblock \href {https://doi.org/10.48550/ARXIV.2310.05209} {Scaling laws of rope-based extrapolation}.
\newblock \emph{CoRR}, abs/2310.05209.

\bibitem[{Lu et~al.(2024)Lu, Wang, Huang, Wu, Liu, Ouyang, and Bai}]{fit}
Zeyu Lu, Zidong Wang, Di~Huang, Chengyue Wu, Xihui Liu, Wanli Ouyang, and Lei Bai. 2024.
\newblock \href {https://doi.org/10.48550/ARXIV.2402.12376} {Fit: Flexible vision transformer for diffusion model}.
\newblock \emph{CoRR}, abs/2402.12376.

\bibitem[{Pal et~al.(2023)Pal, Karkhanis, Roberts, Dooley, Sundararajan, and Naidu}]{giraffe}
Arka Pal, Deep Karkhanis, Manley Roberts, Samuel Dooley, Arvind Sundararajan, and Siddartha Naidu. 2023.
\newblock \href {https://doi.org/10.48550/ARXIV.2308.10882} {Giraffe: Adventures in expanding context lengths in llms}.
\newblock \emph{CoRR}, abs/2308.10882.

\bibitem[{Peng et~al.(2023)Peng, Quesnelle, Fan, and Shippole}]{yarn}
Bowen Peng, Jeffrey Quesnelle, Honglu Fan, and Enrico Shippole. 2023.
\newblock \href {https://doi.org/10.48550/ARXIV.2309.00071} {Yarn: Efficient context window extension of large language models}.
\newblock \emph{CoRR}, abs/2309.00071.

\bibitem[{Rozi{\`e}re et~al.(2023)Rozi{\`e}re, Gehring, Gloeckle, Sootla, Gat, Tan, Adi, Liu, Remez, Rapin et~al.}]{roziere2023code}
Baptiste Rozi{\`e}re, Jonas Gehring, Fabian Gloeckle, Sten Sootla, Itai Gat, Xiaoqing~Ellen Tan, Yossi Adi, Jingyu Liu, Tal Remez, J{\'e}r{\'e}my Rapin, et~al. 2023.
\newblock Code llama: Open foundation models for code.
\newblock \emph{arXiv preprint arXiv:2308.12950}.

\bibitem[{Su(2023)}]{rerope}
Jianlin Su. 2023.
\newblock Rectified rotary position embeddings.
\newblock \url{https://github.com/bojone/rerope}.

\bibitem[{Su et~al.(2024)Su, Ahmed, Lu, Pan, Bo, and Liu}]{su2021roformer}
Jianlin Su, Murtadha H.~M. Ahmed, Yu~Lu, Shengfeng Pan, Wen Bo, and Yunfeng Liu. 2024.
\newblock \href {https://doi.org/10.1016/J.NEUCOM.2023.127063} {Roformer: Enhanced transformer with rotary position embedding}.
\newblock \emph{Neurocomputing}, 568:127063.

\bibitem[{Touvron et~al.(2023{\natexlab{a}})Touvron, Lavril, Izacard, Martinet, Lachaux, Lacroix, Rozi{\`{e}}re, Goyal, Hambro, Azhar, Rodriguez, Joulin, Grave, and Lample}]{Llama}
Hugo Touvron, Thibaut Lavril, Gautier Izacard, Xavier Martinet, Marie{-}Anne Lachaux, Timoth{\'{e}}e Lacroix, Baptiste Rozi{\`{e}}re, Naman Goyal, Eric Hambro, Faisal Azhar, Aur{\'{e}}lien Rodriguez, Armand Joulin, Edouard Grave, and Guillaume Lample. 2023{\natexlab{a}}.
\newblock \href {https://doi.org/10.48550/ARXIV.2302.13971} {Llama: Open and efficient foundation language models}.
\newblock \emph{CoRR}, abs/2302.13971.

\bibitem[{Touvron et~al.(2023{\natexlab{b}})Touvron, Martin, Stone, Albert, Almahairi, and et~al.}]{llama2}
Hugo Touvron, Louis Martin, Kevin Stone, Peter Albert, Amjad Almahairi, and et~al. 2023{\natexlab{b}}.
\newblock \href {https://doi.org/10.48550/ARXIV.2307.09288} {Llama 2: Open foundation and fine-tuned chat models}.
\newblock \emph{CoRR}, abs/2307.09288.

\bibitem[{Xia et~al.(2023)Xia, Gao, Zeng, and Chen}]{shearedllama}
Mengzhou Xia, Tianyu Gao, Zhiyuan Zeng, and Danqi Chen. 2023.
\newblock \href {https://doi.org/10.48550/ARXIV.2310.06694} {Sheared llama: Accelerating language model pre-training via structured pruning}.
\newblock \emph{CoRR}, abs/2310.06694.

\bibitem[{Xiao et~al.(2023)Xiao, Tian, Chen, Han, and Lewis}]{attnsink}
Guangxuan Xiao, Yuandong Tian, Beidi Chen, Song Han, and Mike Lewis. 2023.
\newblock \href {https://doi.org/10.48550/ARXIV.2309.17453} {Efficient streaming language models with attention sinks}.
\newblock \emph{CoRR}, abs/2309.17453.

\bibitem[{Xiong et~al.(2023)Xiong, Liu, Molybog, Zhang, Bhargava, Hou, Martin, Rungta, Sankararaman, Oguz, Khabsa, Fang, Mehdad, Narang, Malik, Fan, Bhosale, Edunov, Lewis, Wang, and Ma}]{longcontextscale}
Wenhan Xiong, Jingyu Liu, Igor Molybog, Hejia Zhang, Prajjwal Bhargava, Rui Hou, Louis Martin, Rashi Rungta, Karthik~Abinav Sankararaman, Barlas Oguz, Madian Khabsa, Han Fang, Yashar Mehdad, Sharan Narang, Kshitiz Malik, Angela Fan, Shruti Bhosale, Sergey Edunov, Mike Lewis, Sinong Wang, and Hao Ma. 2023.
\newblock \href {https://doi.org/10.48550/ARXIV.2309.16039} {Effective long-context scaling of foundation models}.
\newblock \emph{CoRR}, abs/2309.16039.

\bibitem[{Zhang et~al.(2023)Zhang, Li, Jin, and Li}]{hit}
Kechi Zhang, Zhuo Li, Zhi Jin, and Ge~Li. 2023.
\newblock \href {https://doi.org/10.1109/ICPC58990.2023.00030} {Implant global and local hierarchy information to sequence based code representation models}.
\newblock In \emph{31st {IEEE/ACM} International Conference on Program Comprehension, {ICPC} 2023, Melbourne, Australia, May 15-16, 2023}, pages 157--168.

\bibitem[{Zhang et~al.(2024)Zhang, Zeng, Wang, and Lu}]{tinyllama}
Peiyuan Zhang, Guangtao Zeng, Tianduo Wang, and Wei Lu. 2024.
\newblock \href {https://doi.org/10.48550/ARXIV.2401.02385} {Tinyllama: An open-source small language model}.
\newblock \emph{CoRR}, abs/2401.02385.

\end{thebibliography}

% \clearpage
\clearpage
\newpage
\appendix

% \section{Details of Code Symbol Understanding task}

% We extract long code files from popular open-sourced code repositories, especially those newly updated code projects to avoid data leakage.
% We construct a static analysis tool to get the abstract syntax tree of each code file, and then get all defined function and class names in it as the ground-truth output symbols.
% Details of these code projects are shown in Table \ref{tab:codeprojectstat}. We show the number of files we extracted from each project as well as the average length of these code files.
% We also show the average number of symbols and their location statistics in the table.

% In Figure \ref{fig:newtask} we show an illustration of our proposed new task. We replace the \textit{input\_code} part with each code file, and use the task prompt to guide models to extract and output all defined function and class names in input code. We also show an example output in the figure.

% \input{table/codeprojectstat}
% \begin{figure}[!h]
% \centering
%   \includegraphics[width=0.7\columnwidth]{figure/newtask.pdf}  
% \caption{Illustration of Code Symbol Understanding task. We use the task prompt to guide models to extract and output all defined function and class names in input code.}
% \label{fig:newtask}
% % \vspace{-18pt}
% \end{figure}

\section{Details of ShortLLM experiments}

\begin{figure*}[!t]
\centering
  \includegraphics[width=0.7\linewidth]{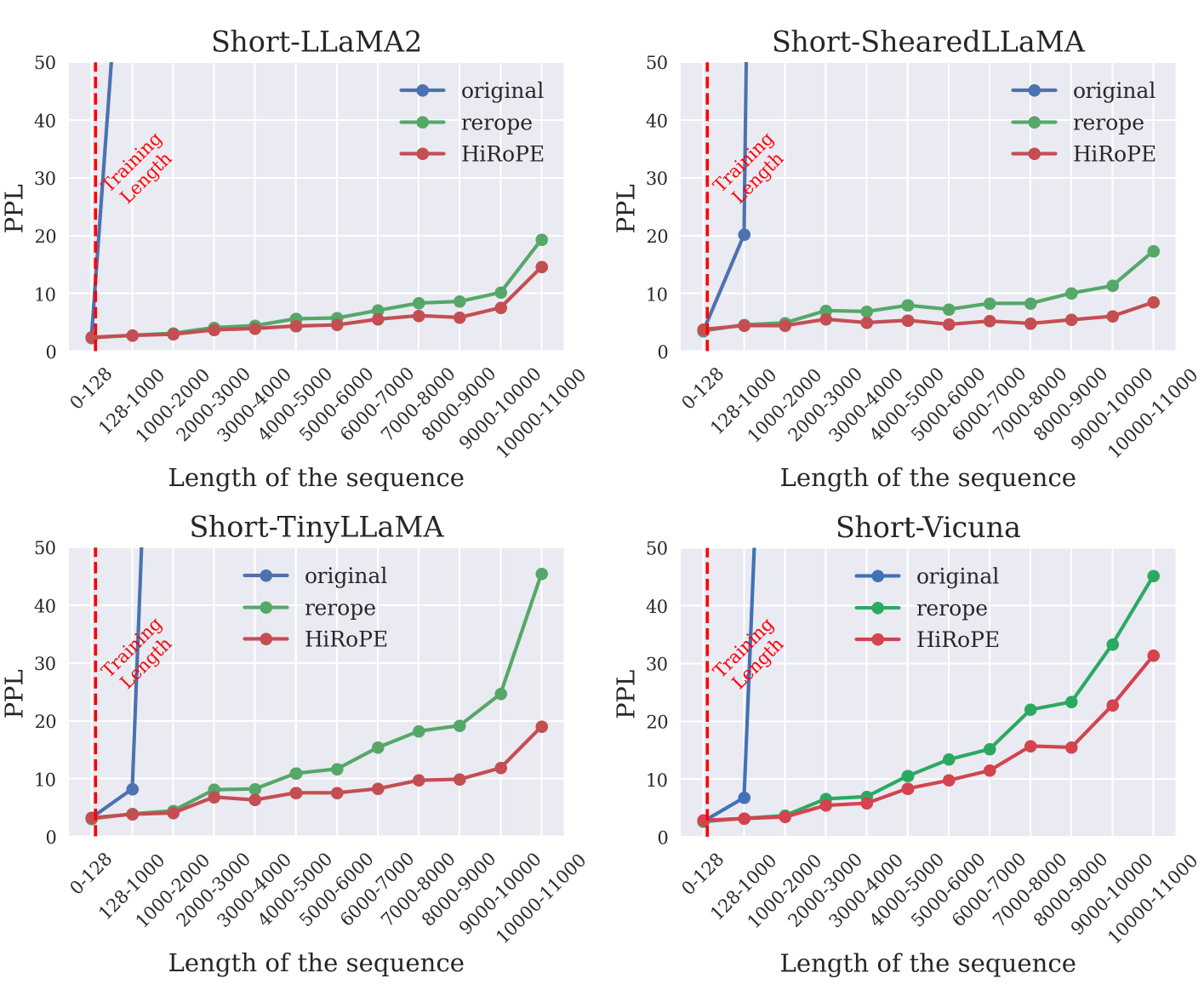}  
\caption{Performance of ShortLLMs on CodeParrot dataset. The training length is set to 128.}
\label{fig:shortenllm}
\end{figure*}

\subsection{Theoretical calculation for O.O.D issues}

According to the analysis in Section \ref{sec:ood}, we can calculate the reliable dimensions for each model as:
\begin{equation}
\small
\begin{split}
    Split\% = log_{10000} \frac{L_{pretrain}}{2\pi}
\end{split}
\label{eq:splitood}
\end{equation}
\begin{table}[!h] 
\footnotesize
    \centering
    \resizebox{\linewidth}{!}{
\centering
% Please add the following required packages to your document preamble:
% \usepackage{multirow}

\begin{tabular}{l|ccc|cc}
\toprule
             & \multicolumn{1}{l}{$L_{pretrain}$} & \multicolumn{1}{l}{RoPE Dim} &  & \multicolumn{1}{l}{Split \%} & \multicolumn{1}{l}{Split Dim} \\
\midrule
LLaMA-2      & 4096                             & 128                          &                      & 0.70                        & 90.05                         \\
% \midrule
ShearedLLaMA & 4096                             & 128                          &                      & 0.70                        & 90.05                         \\
% \midrule
TinyLLaMA    & 2048                             & 64                           &                      & 0.63                        & 40.21                         \\
% \midrule
Vicuna       & 2048                             & 128                          &                      & 0.63                        & 80.42                       \\
\bottomrule
\end{tabular}

}
\caption{Theoretical calculation of the reliable extrapolation dimension.}
   \label{tab:OOD}
\end{table}

The theoretical calculation results are shown in Table \ref{tab:OOD}.

\subsection{ShortLLM Training}

\begin{table}[!t] 
\footnotesize
    \centering
    \resizebox{\linewidth}{!}{
\centering
% Please add the following required packages to your document preamble:
% \usepackage{multirow}

\begin{tabular}{l|ccc|cc}
\toprule
             & Para. & $L_{pretrain}$ &  & $L_{short}$ & $\alpha_{short}$ \\
             \midrule
LLaMA-2      & 7B    & 4096         &  & 128    & 32     \\
ShearedLLaMA & 1.3B  & 4096         &  & 128    & 32     \\
TinyLLaMA    & 1.1B  & 2048         &  & 128    & 16     \\
Vicuna       & 7B    & 2048         &  & 128    & 16     \\
% CodeLLaMA    & 7B    & 16384        &  & 128    & 128   \\
\bottomrule
\end{tabular}

}
\caption{Model Statistics for ShortLLM experiments.}
   \label{tab:shortllmstat}
\end{table}

In order to obtain the ShortLLM of suitable length, we use the position interpolation \cite{PI} method \textbf{\textit{reversely}}. 
Given a position index $m$ in the original RoPE, we use the new index $\alpha_{short} * m$ to replace it as shown in Table \ref{tab:shortllmstat}. 
We sample the CodeParrot-train dataset and filter the data length less than $L_{short}$. We set the global batch size to 64 and fine-tune the models for 1000 steps.

\subsection{ShortLLM Performances}

After training, we apply our HiRoPE to these ShortLLM models. Figure \ref{fig:shortenllm} shows the experiment results of the ShortLLM.
% Our trained ShortLLM successfully demonstrates the expected performance: the performance decreases after surpassing the training length.
We can observe that for all models, Our HiRoPE can maintain good stability as the length of the input code increases. Even if the baseline ReRoPE method gradually becomes unstable under some experimental settings (such as on Short-TinyLLaMA), our method can resist these performance declines.

\end{document}